\documentclass[twocolumn]{article}

\usepackage{cite}
\usepackage{amsmath,amssymb,amsfonts}
\usepackage{algorithmic}
\usepackage{graphicx}
\usepackage{textcomp}
\usepackage{xcolor}

\usepackage{multirow}
\usepackage{multicol}
\usepackage{xspace}

\usepackage[boxed,ruled]{algorithm2e}
\usepackage{graphicx}
\usepackage{makecell}
\usepackage{amsmath}
\usepackage{url}

\def\BibTeX{{\rm B\kern-.05em{\sc i\kern-.025em b}\kern-.08em
    T\kern-.1667em\lower.7ex\hbox{E}\kern-.125emX}}

\advance\topmargin-1in\advance\oddsidemargin-1in
\evensidemargin\oddsidemargin

\usepackage{sectsty}

\sectionfont{\fontsize{12}{15}\selectfont}
\subsectionfont{\fontsize{10}{15}\selectfont}
\subsubsectionfont{\fontsize{9}{15}\selectfont}

\usepackage[left=1.6cm,right=1.6cm,top=1.5cm,bottom=1.5cm]{geometry}

\usepackage{etoolbox}
\makeatletter
\patchcmd{\maketitle}
 {\def\@makefnmark}
 {\def\@makefnmark{}\def\useless@macro}
 {}{}
\makeatother

\begin{document}
\date{}

\title{\Large\textbf{Tiny but Accurate: A Pruned, Quantized and Optimized Memristor Crossbar Framework for Ultra Efficient DNN Implementation}\vspace{-0.3cm}}

\author{\small
Xiaolong Ma\text{$^\dagger$}\textsuperscript{1}, Geng Yuan\text{$^\dagger$}\textsuperscript{1}\thanks{$^\dagger$These authors contributed equally.}, Sheng Lin\textsuperscript{1}, Caiwen Ding\textsuperscript{2}, Fuxun Yu\textsuperscript{3}, Tao Liu\textsuperscript{4}, Wujie Wen\textsuperscript{4}, Xiang Chen\textsuperscript{3}, Yanzhi Wang\textsuperscript{1}\\
\small\textsuperscript{1}Northeastern University, 
\small\textsuperscript{2}University of Connecticut, 
\small\textsuperscript{3}George Mason University, 
\small\textsuperscript{4}Florida International University  \\
\small{E-mail:} \small\textsuperscript{1}\{ma.xiaol, yuan.geng, lin.sheng,\}@husky.neu.edu, \textsuperscript{1}yanz.wang@northeastern.edu,\\
\small\textsuperscript{2}caiwen.ding@uconn.edu, \textsuperscript{3}\{fyu2, xchen26\}@gmu.edu, \textsuperscript{4}\{tliu023, wwen\}@fiu.edu}

\maketitle
\thispagestyle{empty}

{\small\textbf{Abstract---
The state-of-art DNN structures involve intensive computation and high memory storage.  To mitigate the challenges, the memristor crossbar array has emerged as an intrinsically suitable matrix computation and low-power acceleration framework for DNN applications. 
However, the high accuracy solution for extreme model compression on memristor crossbar array architecture is still waiting for unraveling.
In this paper, we propose a memristor-based DNN framework which combines both structured weight pruning and quantization by incorporating \textit{alternating direction method of multipliers} (ADMM) algorithm for better pruning and quantization performance. We also discover the non-optimality of the ADMM solution in weight pruning and the unused data path in a structured pruned model.
Motivated by these discoveries, we design a software-hardware co-optimization framework which contains the first proposed \textit{Network Purification} and \textit{Unused Path Removal} algorithms targeting on post-processing a structured pruned model after ADMM steps. By taking memristor hardware constraints into our whole framework, we achieve extreme high compression ratio on the state-of-art neural network structures with minimum accuracy loss. For quantizing structured pruned model, our framework achieves nearly no accuracy loss after quantizing weights to 8-bit memristor weight representation. We share our models at anonymous link \textcolor{blue}{\url{https://bit.ly/2VnMUy0}}.
}}

\vspace{-3mm}
\section{Introduction}

Structured weight pruning~\cite{wen2016learning,ma2019resnet,zhang2018adam} and weight quantization~\cite{park2017weighted,wu2016quantized,lin2019toward} techniques are developed to facilitate weight compression and computation acceleration to solve the high demand for parallel computation and storage resources~\cite{niu201926ms,li2019admm,ding2018structured}. Even though models are compressed, computation complexity still burden the overall performance of the state-of-art CMOS hardware applications. 


To mitigate the bottleneck caused by CMOS-based DNN architectures, the next-generation device/circuit technologies~\cite{strukov2008missing,ma2018area} triumph CMOS in their non-volatility, high energy efficiency, in-memory computing capability and high scalability. Memristor crossbar device has shown its potential for bearing all these characteristic which makes it intrinsically suitable for large DNN hardware architecture design. A memristor crossbar device can perform matrix-vector multiplication in the analog domain and the computation is in $O(1)$ time complexity~\cite{chua1971memristor,yuan2017memristor}. 
Motivated by the fact that there is no precedent model that is structured pruned and quantized as well as satisfying memristor hardware constraints, in this work, a \textit{memristor-based ADMM regularized optimization} method is utilized both on structured pruning and weight quantization in order to mitigate the accuracy degradation during extreme model compression. A structured pruned model can potentially benefit for high-parallelism implementation in crossbar architecture. Further more, quantized weights can reduce hardware imprecision during read/write procedure, and save more hardware footprint due to less peripheral circuits are needed to support fewer bits.

However, to achieve ultra-high compression ratio, an ADMM pruning method~\cite{zhang2018adam,ye2019progressive} cannot fully exploit all redundancy in a neural network model. As a result, we design a hardware-software co-optimization framework in which we investigate \textit{Network Purification} and \textit{Unused Path Removal} after the procedure of \textit{structured weight pruning with ADMM}. Moreover, we utilize distilled knowledge from software model to guide our memristor hardware constraint quantization. To the best of our knowledge, we are the first to combine extreme structured weight pruning and weight quantization in an unified and systematic memristor-based framework. Also, we are the first to discover the redundant weights and unused path in a structured pruned DNN model and design a sophisticate co-optimization framework to boost higher model compression rate as well as maintain high network accuracy. By incorporating memristor hardware constraints in our model, our frameworks are guaranteed feasible for a real memristor crossbar device. The contributions of this paper include:

\begin{itemize}
    \item We adopt ADMM for efficiently optimizing the non-convex problem and utilized this method on structured weight pruning.
    \item We systematically investigate the weight quantization on a pruned model with memristor hardware constraints.
    \item We design a software-hardware co-optimization framework in which \textit{Network Purification} and \textit{Unused Path Removal} are first proposed.
    
\end{itemize}

We evaluate our proposed memristor framework on different networks. We conclude that structured pruning method with \textit{memristor-based ADMM regularized optimization} achieves high compression ratio and desirable high accuracy. 
Software and hardware experimental results shows our memristor framework is very energy efficient and saves great amount of hardware footprint. 

\vspace{-4mm}
\section{Related Works}

Heuristic weight pruning methods~\cite{han2015learning} are widely used in neuromorphic computing designs to reduce the weight storage and computing delay~\cite{ankit2017trannsformer}. 
\cite{ankit2017trannsformer} implemented weight pruning techniques on a neuromorphic computing system using irregular pruning caused unbalanced workload, greater circuits overheads and extra memory requirement on indices.
To overcome the limitations, \cite{wang2017group} proposed group connection deletion, which structually prunes connections to reduce routing congestion between memristor crossbar arrays. 

Weight quantization can mitigate hardware imperfection of memristor including state drift and process variations, caused by the imperfect fabrication process or by the device feature itself~\cite{park2017weighted,wu2016quantized}. 
\cite{xia2016switched} presented a technique to reduce the overhead of Digital-to-Analog Converters (DACs)/Analog-to-Digital Converters (ADCs) in resistive random-access memory (ReRAM) neuromorphic computing systems. They first normalized the data, and then quantized intermediary data to 1-bit value. This can be directly used as the analog input for ReRAM crossbar and, hence, avoids the need of DACs.

\vspace{-4mm}
\section{Background on Memristors}
\label{sec:2}

\vspace{-3mm}
\subsection{Memristor Crossbar Model}
\begin{figure} [t]
     \centering
     \includegraphics[width=0.7\columnwidth]{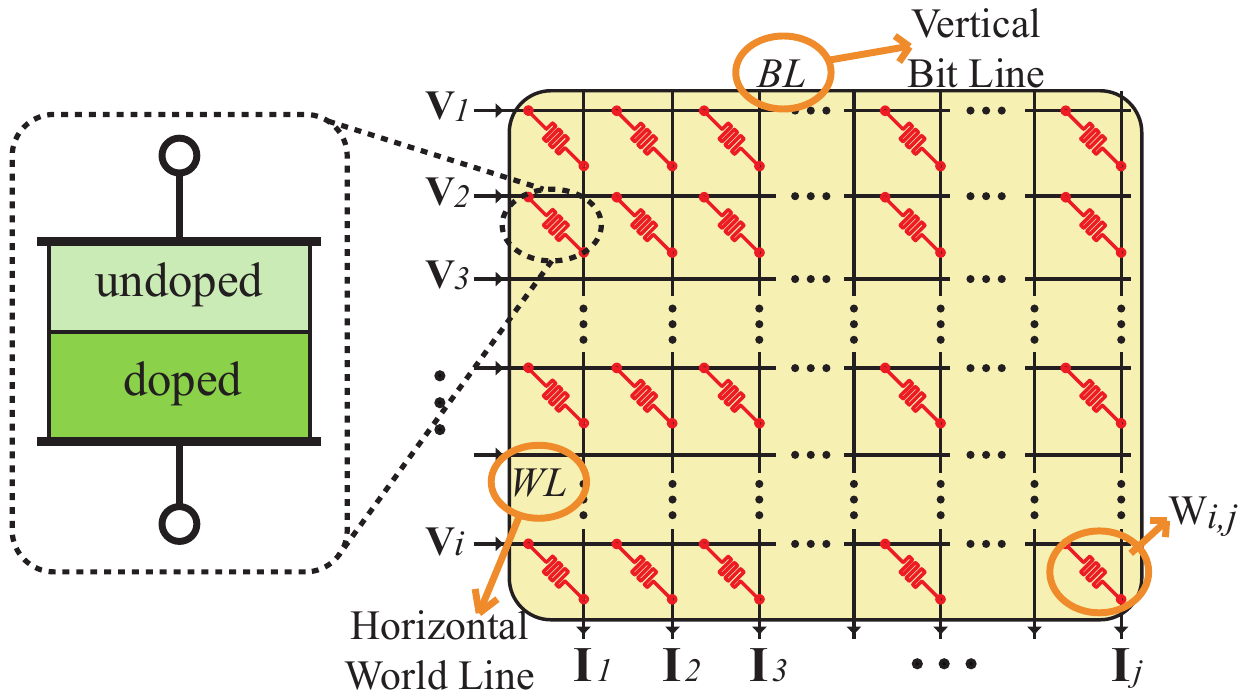}  
     \vspace{-3mm}
     \caption{memristor and memristor crossbar}
     \label{fig:crossbar}
 \end{figure}

Memristor~\cite{strukov2008missing} crossbar is an array structure consists of memristors, horizontal Word-lines and Vertical Bit-lines, as shown in Figure~\ref{fig:crossbar}. Due to its outstanding performance on computing matrix-vector multiplications (MVM), memristor crossbars are widely used as dot-product accelerator in recent neuromorphic computing designs~\cite{Shafiee2016}. By programming the conductance state (which is also known as ``memristance'') of each memristor, the weight matrix $\bf{W}$ can be mapped onto the memristor crossbar. Given the input voltage vector $\textbf{V}_i$, the MVM output current vector $\textbf{I}_j$ can be obtained in time complexity of $O(1)$.

\vspace{-4mm}
\subsection{Challenges in Memristor Crossbars Implementation and Mitigation Techniques}

Different from the software-based designs, hardware imperfection is one of the key issues that causes the hardware non-ideal behaviors and needs to be considered in memristor-based designs. The hardware imperfection of memristor devices are mainly come from the imperfect fabrication process and the memristor features. 

\emph{\textbf{Process Variation.}} Process variation is one major hardware imperfection that caused by the fluctuations in fabrication process. It mainly comes from the line-edge roughness, oxide thickness fluctuations, and random dopant variations~\cite{ roughness}. Inevitably, process variation plays an increasingly significant role as the process technology scales down to nanometer level. In a DNN hardware design, the non-ideal behaviors caused by process variations may lead to an accuracy degradation.

\emph{\textbf{State Drift.}} State drift is the phenomenon that the memristance would change after several reading opertions~\cite{Yang2008}. It is known that memristor is a thin-film device constructed by a region highly doped with oxygen vacancies and an undoped region. By nature, applying an electric field across the memristor over a period of time, the oxygen vacancies would migrate to the direction along with the electric field, which leads to the (memristance) state  drift. Consequently, an error will incur when the state of memristor drifts to another state level.

It has been proved that applying quantization on memristor-based designs can mitigate the undesired impacts caused by hardware imperfections~\cite{song2017quantization}.
\vspace{-5mm}
\section{A Memristor-Based Highly Compressed DNN Framework}


The memristor crossbar structure has shown its potential for neuromorphic computing system compared to the CMOS technologies\cite{ankit2017trannsformer}. Due to great amount of weights and computations that involved in networks, an efficient and highly performed framework is needed to conquer the memory storage and energy consumption problems. We propose an unified memristor-based framework including \textit{memristor-based ADMM regularized  optimization} and \textit{masked mapping}. 


\vspace{-4mm}
\subsection{Problem Formulation}

ADMM\cite{boyd2011distributed} is an advanced optimization technique which decompose an original problem into subproblems that can be solved separately and iteratively. 
By adopting \textit{memristor-based ADMM regularized  optimization}, the framework can guarantee the solution feasibility (satisfying memristor hardware constraints) while provide high solution quality (no obvious accuracy degradation after pruning).



First, the {\textit{memristor-based ADMM regularized  optimization}} starts from a pre-trained full size DNN model without compression. Consider an $N$-layer DNNs, sets of weights of the $i$-th (CONV or FC) layer are denoted by ${\bf{W}}_{i}$. And the \textit{loss function} associated with the DNN is denoted by $f \big( \{{\bf{W}}_{i}\}_{i=1}^N \big)$. The overall problem is defined by
\begin{equation}
\small
\label{original}
\begin{aligned}
& \underset{ \{{\bf{W}}_{i}\}}{\text{minimize}}
& & f \big( \{{\bf{W}}_{i}\}_{i=1}^N \big),
\\ & \text{subject to}
& & {\bf{W}}_{i}\in {\bf{\mathcal{P}}}_{i}, \; {\bf{W}}_{i}\in {\bf{\mathcal{Q}}}_{i}, \; i = 1, \ldots, N.
\end{aligned}
\end{equation}
Given the value of $\alpha_{i}$, the memristor-based constraint set 
${{\bf{\mathcal{P}}}_{i}=\{{\bf{W}}_{i}|\sum(\text{structured }\bf{W}}_{i}\neq 0)\leq\alpha_{i}\}$ 
and ${\mathcal{Q}}_{i}$=\{the weights in the $i$-th layer are mapped to the quantization values\},
where $\alpha_i$ is predefined hyper parameters. The general constraint can be extended in structured pruning such as filter pruning, channel pruning and column pruning, which facilitate high-parallelism implementation in hardware. 

Similarly, for weight quantization, elements in ${\bf{\mathcal{Q}}}_{i}$ are the solutions of ${\bf{W}}_{i}$. Assume set of ${q_{i,1}, q_{i,2}, \cdots, q_{i,M_{i}}}$ is the available memristor state value which is the elements in ${\bf{W}}_{i}$, where $M_i$ denotes the number of available quantization level in layer $i$. Suppose $q_{i,j}$ indicates the $j$-th quantization level in layer $i$, which gives
\begin{equation}
\footnotesize
    q_{i,j}\in[-memr_{max},-memr_{min}] \cup [memr_{min},memr_{max}]
\end{equation}
where $memr_{min},$ $memr_{max}$ are the minimum and maximum memristance value of a specified memristor device.

\begin{figure} [t]
    \centering
    \includegraphics[width=0.7\columnwidth]{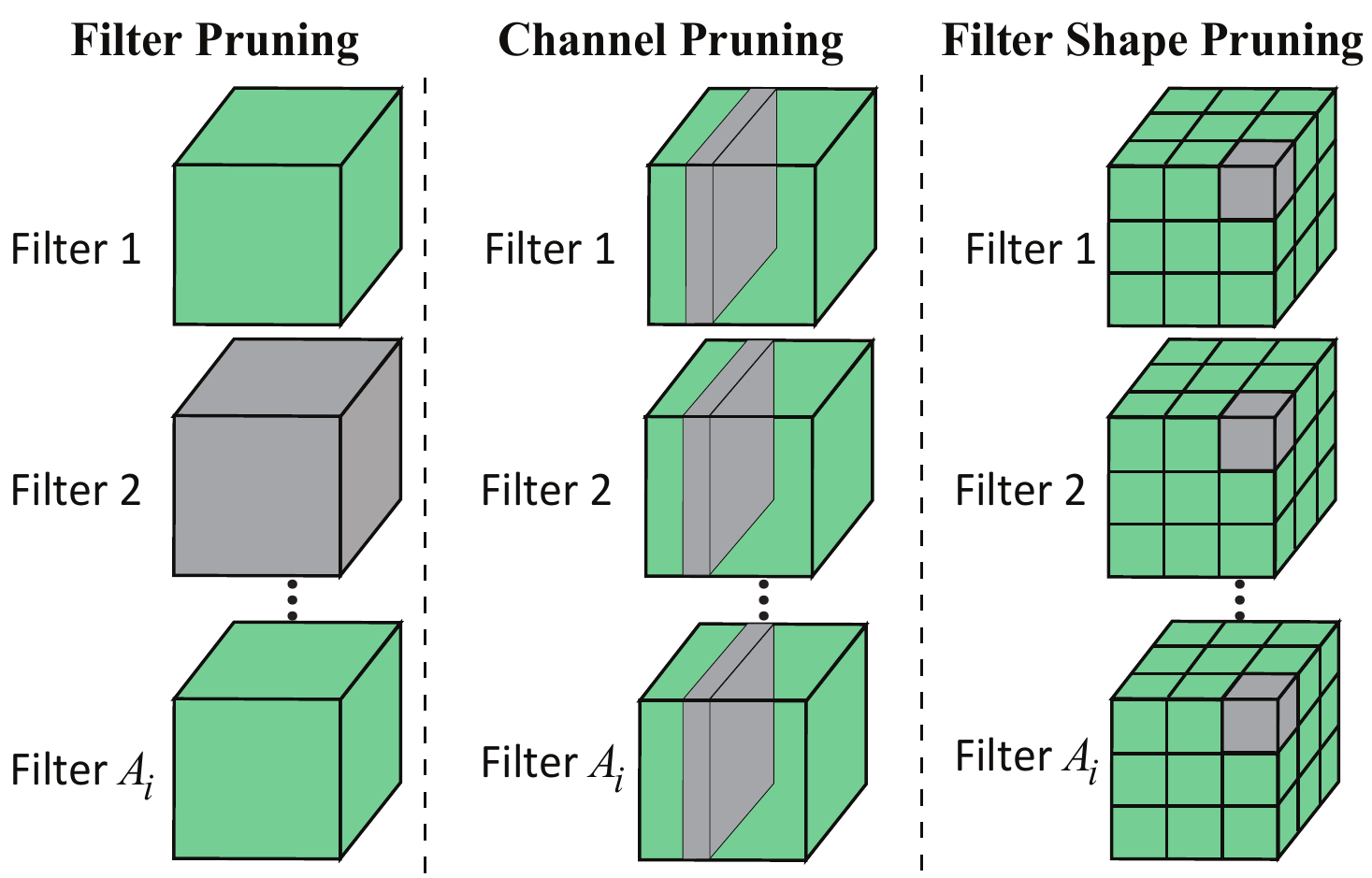}     
    \vspace{-3mm}
    \caption{Illustration of filter-wise, channel-wise and shape-wise structured sparsities.}
    \label{fig:pruning}
\end{figure}

\begin{figure*} [t]
     \centering
     \includegraphics[width=1\textwidth]{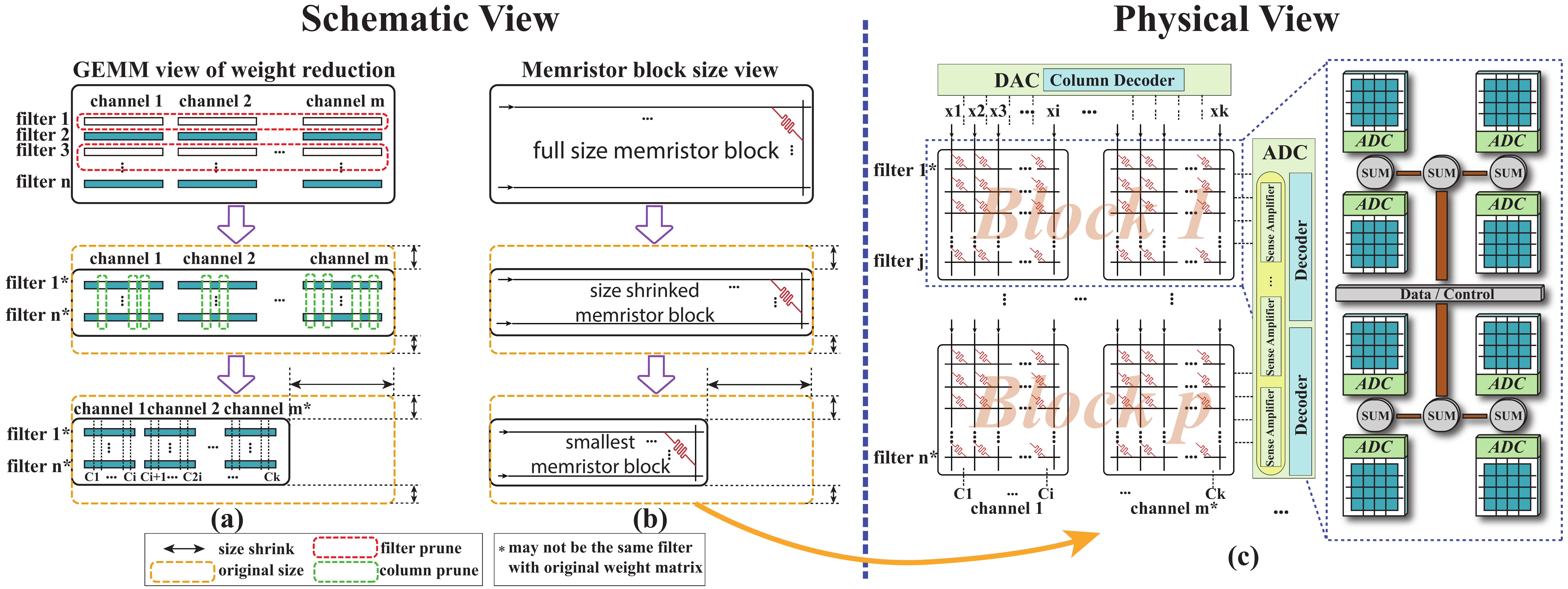}   
     \vspace{-5mm}
     \caption{Structured weight pruning and reduction of hardware resources}
     \label{fig:gemm_memristor_block}
\end{figure*}
\vspace{-3mm}
\subsection{Memristor-based ADMM regularized optimization step}
Corresponding to every memristor-based constraint set of $\bf{\mathcal{P}_{i}}$ and $\bf{\mathcal{Q}_{i}}$, 
a indicator functions is utilized to incorporate $\bf{\mathcal{P}_{i}}$ and $\bf{\mathcal{Q}_{i}}$ into objective functions, which are

\begin{minipage}{0.48\linewidth}
\small
\begin{eqnarray*}g_{i}({\bf{W}}_{i})=
\begin{cases}
 0 & \text { if } {\bf{W}}_{i}\in {\bf{\mathcal{P}}}_{i}, \\ 
 +\infty & \text { otherwise, }
\end{cases}
\end{eqnarray*}
\end{minipage}
\begin{minipage}{0.47\linewidth}
\small
\begin{eqnarray*}h_{i}({\bf{W}}_{i})=
\begin{cases}
 0 & \text { if } {\bf{W}}_{i}\in {\bf{\mathcal{Q}}}_{i}, \\ 
 +\infty & \text { otherwise, }
\end{cases}
\end{eqnarray*}
\end{minipage}

for $i = 1, \ldots, N$. Substituting into (\ref{original}) and we get 
\begin{equation}
\small
\label{admm_form}
\begin{aligned}
& \underset{ \{{\bf{W}}_{i}\}}{\text{minimize}}
& & f \big( \{{\bf{W}}_{i} \}_{i=1}^N \big)+\sum_{i=1}^{N} g_{i}({\bf{Y}}_{i})+\sum_{i=1}^{N} h_{i}({\bf{Z}}_{i}),
\\ & \text{subject to}
& & {\bf{W}}_{i} = \textbf{Y}_i = \textbf{Z}_i, \; i = 1, \ldots, N,
\end{aligned}
\end{equation}

We incorporate auxiliary variables ${\bf{Y}}_{i}$ and ${\bf{Z}}_{i}$, dual variables ${\bf{U}}_{i}$ and ${\bf{V}}_{i}$, and the augmented Lagrangian formation $L_\rho{\{\cdot\}}$ of problem (\ref{admm_form}) is
\vspace{-0.50em}
\begin{equation}
\small
\begin{aligned}
\label{equ7}
 \underset{ \{{\bf{W}}_{i}\}}{\text{minimize}}
\ \ \ & f \big( \{{\bf{W}}_{i} \}_{i=1}^N \big) +\sum_{i=1}^{N} \frac{\rho_{i}}{2}  \| {\bf{W}}_{i}-{\bf{Y}}_{i}+{\bf{U}}_{i} \|_{F}^{2}\\ & + \sum_{i=1}^{N} \frac{\rho_{i}}{2}  \| {\bf{W}}_{i}-{\bf{Z}}_{i}+{\bf{V}}_{i} \|_{F}^{2}, \\
\end{aligned}
\end{equation}

We can see that the first term in problem (\ref{equ7}) is original DNN loss function, and the second and third term are differentiable and convex. As a result, subproblem (\ref{equ7}) can be solved by stochastic gradient descent~\cite{kingma2014adam} as the original DNN training.

The standard ADMM algorithm~\cite{boyd2011distributed} steps proceed by repeating, for $k = 0, 1,\ldots$, the following subproblems iterations:
\begin{equation}
\small
\label{itera1}
\begin{aligned}
 \bf{W}_{i}^{k+1}:=\underset{ \{{\bf{W}}_{i}\}}{\text{minimize}}& \ \textit{L}_\rho(\{\bf{W}_i\}, \{\bf{Y}_i^k\}, \{\bf{U}_i^k\})\\
& + \textit{L}_\rho(\{\bf{W}_i\}, \{\bf{Z}_i^k\}, \{\bf{V}_i^k\})
\end{aligned}
\end{equation}
\begin{equation}
\small
\label{itera2}
\begin{aligned}
 \bf{Y}_{i}^{k+1}, \bf{Z}_{i}^{k+1}:=\underset{ \{{\bf{Y}}_{i}, \bf{Z}_{i}\}}{\text{minimize}}& \ \textit{L}_\rho(\{\bf{W}_i^{k+1}\}, \{\bf{Y}_i\}, \{\bf{U}_i^k\})\\
& + \textit{L}_\rho(\{\bf{W}_i^{k+1}\}, \{\bf{Z}_i\}, \{\bf{V}_i^k\})
\end{aligned}
\end{equation}
\begin{equation}
\small
    \bf{U}_{i}^{k+1} := \bf{U}_{i}^{k}+\bf{W}_{i}^{k+1}-\bf{Y}_{i}^{k+1};\ \bf{V}_{i}^{k+1} := \bf{V}_{i}^{k}+\bf{W}_{i}^{k+1}-\bf{Z}_{i}^{k+1}
\label{itera3}
\end{equation}

which (\ref{itera1}) is the proximal step, (\ref{itera2}) is projection step and (\ref{itera3}) is dual variables update.

The optimal solution is the Euclidean projection (masked mapping) of ${\bf{W}}_{i}^{k+1}+{\bf{U}}_{i}^{k}$ and ${\bf{W}}_{i}^{k+1}+{\bf{V}}_{i}^{k}$ onto ${\mathcal{P}}_{i}$ and $\bf{\mathcal{Q}_{i}}$. Namely, elements in solution that less than $\alpha_i$ will be set to zero. In the meantime, those kept elements are quantized to the closest valid memristor state value.

\vspace{-4mm}
\subsection{Memristor-Based Structured Weight Pruning} In order to accommodate high-parallelism implementation in hardware, we use structured pruning method~\cite{wen2016learning} instead of the irregular pruning method~\cite{han2015learning} to reduce the size of the weight matrix while avoid extra memory storage requirement for indices. Figure~\ref{fig:pruning} shows different types of structured sparsity which include filter-wise sparsity, channel-wise sparsity and shape-wise sparsity. 

Figure \ref{fig:gemm_memristor_block} (a) shows the general matrix multiplication (GEMM) view of the DNN weight matrix and the different structured weight pruning methods. The structured pruning corresponds to removing rows (filters-wise) or columns (shape-wise) or the combination of them. 
We can see that after structured weight pruning, the remaining weight matrix is still regular and without extra indices.

Figure \ref{fig:gemm_memristor_block} (b) illustrate the memristor crossbar schematic size reduction from corresponding structured weight pruning and Figure~\ref{fig:gemm_memristor_block} (c) shows physical view of the memristor crossbar blocks. A CONV layer has $n$ filters, $m$ channels which include total $k$ columns, and is denoted as ${\bf{W}}\in\mathbb{R}^{n \times k}$. Due to the increasing reading/writing errors caused by expanding the memristor crossbar size, we limited our design by using multiple 128$\times$64~\cite{Hu2018} crossbars for all DNN layers. In Figure~\ref{fig:gemm_memristor_block} (c), $i, j$ denote columns and rows for each crossbar, $X$ represent inputs and $c$ is the column number which is also shown in Figure ~\ref{fig:gemm_memristor_block} (a). By easy calculation, one can derived that there's $k/j$ different crossbars to store one filter's weights as a block unit. So there's total $p=n/j$ blocks to store ${\bf{W}}\in\mathbb{R}^{n \times k}$. Within each block, the outputs of each crossbar will be propagated through an ADC. Then We column-wisely sum the intermediate results of all crossbars.
\vspace{-4mm}
\section{Software-hardware Co-optimization}
Due to the existence of the non-optimality of ADMM process and the accuracy degradation problem of quantizing sparse DNN, a software-hardware co-optimization framework is desired. In this section we propose: (i) network purification and unused path removal to efficiently remove redundant channels or filters, (ii) memristor model quantization by using distilled knowledge from software helper. 

\subsection{Network Purification and Unused Path Removal}
Weight pruning with memristor-based ADMM regularized optimization can significantly reduce the number of weights while maintaining high accuracy. However, does the pruning process really remove all unnecessary weights?

\begin{figure}[h!]
     \centering
     \includegraphics[width=0.98\columnwidth]{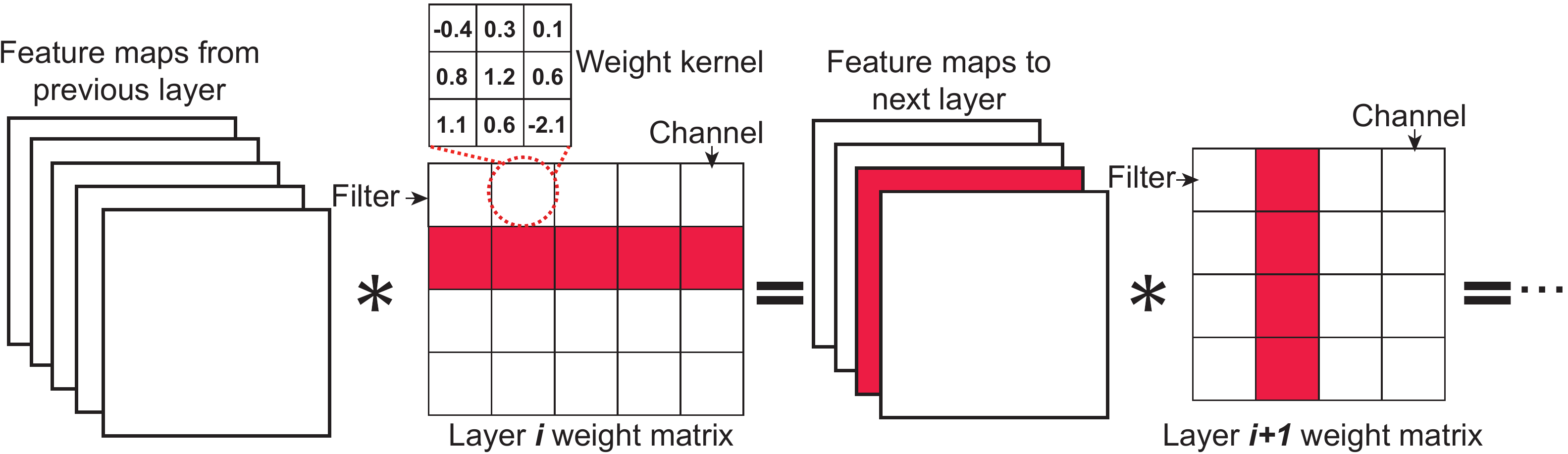}     
     \vspace{-1mm}
     \caption{Unused data path caused by structured pruning}
     \label{fig:unused}
\end{figure}

From our analysis on the DNN data flow, we find that if a whole filter is pruned, after General Matrix Multiply (GEMM), the generated feature maps by this filter will be ``blank". 
If we map those ``blank" feature input to next layer, the corresponding unused input channel weights become removable. 
By the same token, a pruned channel also causes the corresponding filter in previous layer removable. Figure~\ref{fig:unused} gives a clear illustration about the corresponding relationship between pruned filters/channels and correspond unused channels/filters.

To better optimize the unused path removal effect we discussed above, we derive an emptiness ratio parameter $\eta$ to define what can be treated as an empty channel. Suppose $\Lambda_i$ is the number of columns per channel in layer $i$, and $j$ is channel index. We have 
\vspace{-2.0mm}
\begin{equation}
    \eta_{i,j} = \big[\sum_{k=1}^{\delta} (column_k !=0)\big] / \delta \quad \delta \in \Lambda_i
    \label{equ:empty_ratio}
\vspace{-1.5mm}
\end{equation}
If $\eta_{i,j}$ exceeds a pre-defined threshold, we can assume that this channel is empty and thus actually prune every column in it. 
However, if we remove all columns that satisfy $\eta$, dramatic accuracy drop will occur and it will be hard to recover by retraining because some relatively ``important" weights might be removed. To mitigate this problem, we design \textit{Network Purification} algorithm dealing with the \textit{non-optimality} problem of the ADMM process. 
We set-up an criterion constant $\sigma_{i,j}$ to represent channel $j$'s importance score, which is derived from an accumulation procedure:
\vspace{-3mm}
\begin{equation}
\vspace{-2mm}
    \sigma_{i,j} = \sum_{k=1}^{\delta}\ \|column_k\|_F^2 / \delta \quad \delta \in \Lambda_i
    \label{equ:importance_score}
\end{equation}
One can think of this process as if \textit{collection evidence for whether each channel that contains one or several columns need to be removed}. A channel can only be treated as empty when both equation~(\ref{equ:empty_ratio}) and~(\ref{equ:importance_score}) are satisfied. \textit{Network Purification} also works on purifying remaining filters and thus remove more unused path in the network. 
Algorithm \ref{algo_unused} shows our generalized method of the P-RM method where $Th_1 \ldots Th_4$ are hyper-parameter thresholds values.

\begin{table*}[t]
\scriptsize
    \centering
    \caption{Structured weight pruning results on multi-layer network on MNIST, CIFAR-10 and ImageNet datasets. (P-RM: Network Purification and Unused Path Removal). Accuracies in ImageNet results are reported in \emph{\textbf{Top-5}} accuracy.}
    \vspace{-0.1mm}
    \renewcommand{\arraystretch}{0.96}
    \resizebox{\textwidth}{!}{
        \begin{tabular}{|c c c c c c c|} 
					\hline\hline
					\multirow{2}{*}{Method} & \multirow{2}{*}{\makecell{Original model \\ Accuracy}} & \multirow{2}{*}{\makecell{Compression Rate \\ Without P-RM}} & \multirow{2}{*}{\makecell{Accuracy \\ Without P-RM}} & \multirow{2}{*}{\makecell{Prune Ratio \\ With P-RM}} & \multirow{2}{*}{\makecell{Accuracy \\ With P-RM}} &  \multirow{2}{*}{\makecell{Weight Quantization \\Accuracy (8-bit)}}  \\
					&&&&&& \\
					\hline
					\multicolumn{7}{c}{\textbf{MNIST}} \\
					\hline
					 Group Scissor~\cite{wang2017group} & 99.15\% & 4.16$\times$ & 99.14\% & N/A & N/A & N/A \\ 
					\hline
					\multirow{3}{*}{\makecell{\textbf{our} \\ \textbf{LeNet-5}}} & \multirow{3}{*}{99.17\%} & 23.18$\times$ & 99.20\% & \bf{39.23}$\times$  & \bf{99.20}\%  & \bf{99.16\%} \\ 
					 &  & 34.46$\times$ & 99.06\% &  \text{*}\bf{87.93}$\times$   & \bf{99.06}\% & \bf{99.04}\% \\ 
					 &  & 45.54$\times$ & 98.48\% & \bf{231.82}$\times$    & \bf{98.48}\% & \bf{98.05}\% \\ 
					\hline
					\multicolumn{7}{r}{\text{*}numbers of parameter reduced: \textbf{25.2K}}  \\
					\hline
					\multicolumn{7}{c}{\textbf{CIFAR-10}} \\
					\hline
					 Group Scissor~\cite{wang2017group} & 82.01\% & 2.35$\times$ & 82.09\% & N/A & N/A & N/A \\ 
					\hline
					\multirow{3}{*}{\makecell{\textbf{our} \\ \textbf{ConvNet}}} & \multirow{3}{*}{84.41\%} & 2.35$\times$ & 84.55\% & N/A  & N/A & \bf{84.33}\% \\ 
					 &  & \text{*}2.93$\times$ & 84.53\% & N/A & N/A & \bf{83.93}\% \\ 
					 &  & 5.88$\times$ & 83.58\% & N/A & N/A & \bf{83.01}\% \\ 
					\hline
					\multirow{2}{*}{\makecell{\textbf{our} \\ \textbf{VGG-16}}} & \multirow{2}{*}{93.70\%} & \multirow{2}{*}{20.16$\times$} & 
					\multirow{2}{*}{93.36\%} & \bf{44.67}$\times$ & \bf{93.36}\% & \bf{93.04}\%  \\
					 &  &  &  &  \text{*}\bf{50.02}$\times$ & \bf{92.73}\% & \bf{92.46}\% \\ 
					 \hline
					 \multirow{2}{*}{\makecell{\textbf{our} \\ \textbf{ResNet-18}}} & \multirow{2}{*}{94.14\%} & 5.83$\times$ & 93.79\% & \bf{52.07}$\times$  & \bf{93.79}\% & \bf{93.71}\% \\ 
					 &  & 15.14$\times$ & 93.20\% & \text{*}\bf{59.84}$\times$ & \bf{93.22}\% & \bf{93.27}\% \\ 
					 \hline
					 \multicolumn{7}{r}{\text{*}numbers of  parameter reduced on \textit{ConvNet}: \textbf{102.30K}, \textit{VGG-16}: \textbf{14.42M}, \textit{ResNet-18}: \textbf{10.97M}}  \\
					 \hline
					\multicolumn{7}{c}{\textbf{ImageNet ILSVRC-2012}} \\
					\hline
					SSL~\cite{wen2016learning} AlexNet & 80.40\% & 1.40$\times$ & 80.40\% & N/A & N/A & N/A \\
					\hline
					\bf{our AlexNet} & 82.40\% & 4.69$\times$ & 81.76\% & \bf{5.13}$\times$ & \bf{81.76}\% & \bf{80.45}\% \\
					\hline
					\bf{our ResNet-18} & 89.07\% & 3.02$\times$ & 88.41\% & \bf{3.33}$\times$ & \bf{88.36}\% & \bf{88.47\%} \\
					\hline
					\bf{our ResNet-50} & 92.86\% & 2.00$\times$ & 92.26\% & \bf{2.70}$\times$ & \bf{92.27}\% & \bf{92.20\%} \\
					\hline
					\multicolumn{7}{r}{numbers of  parameter reduced on \textit{AlexNet}: \textbf{1.66M}, \textit{ResNet-18}: \textbf{7.81M}, \textit{ResNet-50}: \textbf{14.77M}} \\
					 \hline\hline
					 
        \end{tabular}
    }
    \label{table:results}
    \vspace{-4mm}
\end{table*}

\begin{algorithm}[!h]
\footnotesize
\SetAlgoLined
\KwResult{Redundant weights and unused paths removed}
 Load ADMM pruned model\par
 $\delta$ = numbers of columns per channel\par
 \For{$i \gets 1$ until last layer}{
  \For{$j \gets 1$ until last $channel$ in $layer_i$}{
    \For{\normalfont{\textbf{each: }} $k \in \delta\ \normalfont{\textbf{and}}\ \|column_k\|_F^2 < Th_1$}{
        \normalfont{calculate: } $equation$ (\ref{equ:empty_ratio}), (\ref{equ:importance_score})\;
    }
    \If{$\eta_{i,j} < Th_2$\ \normalfont{\textbf{and}}\ $\sigma_{i,j} < Th_3$}{
        prune($channel_{i,j}$)\par
        prune($filter_{i-1,j}$) \ \bf{when} $i\neq1$\;
    }
    
  }
  \For{$m \gets 1$ until last $filter$ in $layer_i$}{
    \If{$filter_m$ \normalfont{\textbf{is}} empty \normalfont{\textbf{or}} $\|filter_m\|_F^2 < Th_4$}{
        prune($filter_{i,m}$)\par
        prune($channel_{i+1,m}$) \ \bf{when} $i\neq$ \normalfont{last layer index}\;
    }
  }
 }
 \caption{\footnotesize{Network purification \& Unused path removal}}
 \label{algo_unused}
\end{algorithm}
\vspace{-3mm}
\subsection{Memristor Weight Quantization}
Traditionally, DNN in software is composed by 32-bit weights. But on a memristor device, the weights of a neural network are represented by the memristance of the memristor (i.e. the memristance range constraint $\mathcal{Q}_{i}$ in ADMM process). Due to the limited memristance range of the memristor devices, the weight values exceeding memristance range cannot be represented precisely. Meanwhile, the write-on value and the exact value mismatch when mapping weights on memristor crossbar will also cause the reading mismatch if the amount of the value shift exceeds state level range.

In order to mitigate the memristance range limitation and the mapping mismatch, larger range between state level ($q_{i,1}, q_{i,2}, \cdots, q_{i,M_{i}}$) is needed which means fewer bits are representing weights. To better maintain accuracy, we use a pretrained high-accuracy teacher model to provide distillation loss to add on our memristor model (referred as student model) loss to provide better training performance.\vspace{-2mm}
\begin{equation}
\vspace{-1mm}
    l_{student} = (1-\sigma)\mathbb{L}(p_s, p_r) + \sigma\mathcal{T}^2\mathbb{L}(p_s, p_t)
    \label{eqa:distillation}
\end{equation}
The $\mathbb{L}$ in first term in (\ref{eqa:distillation}) is the memristor model (student) loss, and in second term is distillation loss between student and teacher. $p_s$ and $p_t$ are outputs of student and teacher and $p_r$ is the ground-truth label. $\sigma$ is a balancing parameter, and $\mathcal{T}$ is the temperature parameter.
\begin{algorithm}[t]
\footnotesize
\SetAlgoLined
\KwResult{distillation quantization with memristor hardware constraints}
 $student$ $\gets$ model pruned and ready to apply quantization\;
 $teacher$ $\gets$ model with a deeper structure and higher accuracy\;
 \For{$step \gets 1$ until $l_{student}$ converge}{
    $student_q = apply\_quantization(w_s \text{, } \mathcal{Q})$\;
    calculate $\mathcal{T}^2\mathbb{L}(p_s, p_t)$ of $student_q \And teacher$\; 
    back propagate on $student \gets \frac{\partial (\mathcal{T}^2\mathbb{L}(p_s, p_t))}{\partial (student_q)}$\;
 }
 \caption{Distillation Quantization}
 \label{algo_disti}
\end{algorithm}

\section{Experimental Results}\vspace{-3mm}
In this section, we show the experimental results of our proposed memristor-based DNN framework in which structured weight pruning and quantization with memristor-based ADMM regularized optimization are included. Our software-hardware co-optimization framework (i.e. \textit{Network Purification}, \textit{Unused Path Removal} (P-RM)) are also thoroughly compared. We test MNIST dataset on LeNet-5 and CIFAR-10 dataset using ConvNet (4 CONV layers and 1 FC layer), VGG-16 and ResNet-18, and we also show our ImageNet results on AlexNet, ResNet-18 and ResNet-50. The accuracy of pruned and quantized model results are tested based on our software models that incorporated with memristor hardware constraints. Models are trained on an eight NVIDIA GTX-2080Ti GPUs server using PyTorch API. Our memristor model on MATLAB and the NVSim~\cite{Dong2012} is used to calculate power consumption and area cost of the memristors and memristor crossbars. The 1R crossbar structure is used in our design. And we choose the memristor device that has $R_{on}=1M\Omega$ and $R_{off}=10M\Omega$. 
The memristor precision is 4-bit, which indicates that 16 state-levels can be represented by a single memristor device, and two memristors are combined to represent 8-bit weight in our framework. 
For the peripheral circuits, the power and area is calculated based on 45nm technology. And H-tree distribution networks are used to access all the memristor crossbars.
\begin{figure}[t]
     \centering
     \includegraphics[width=0.46\textwidth]{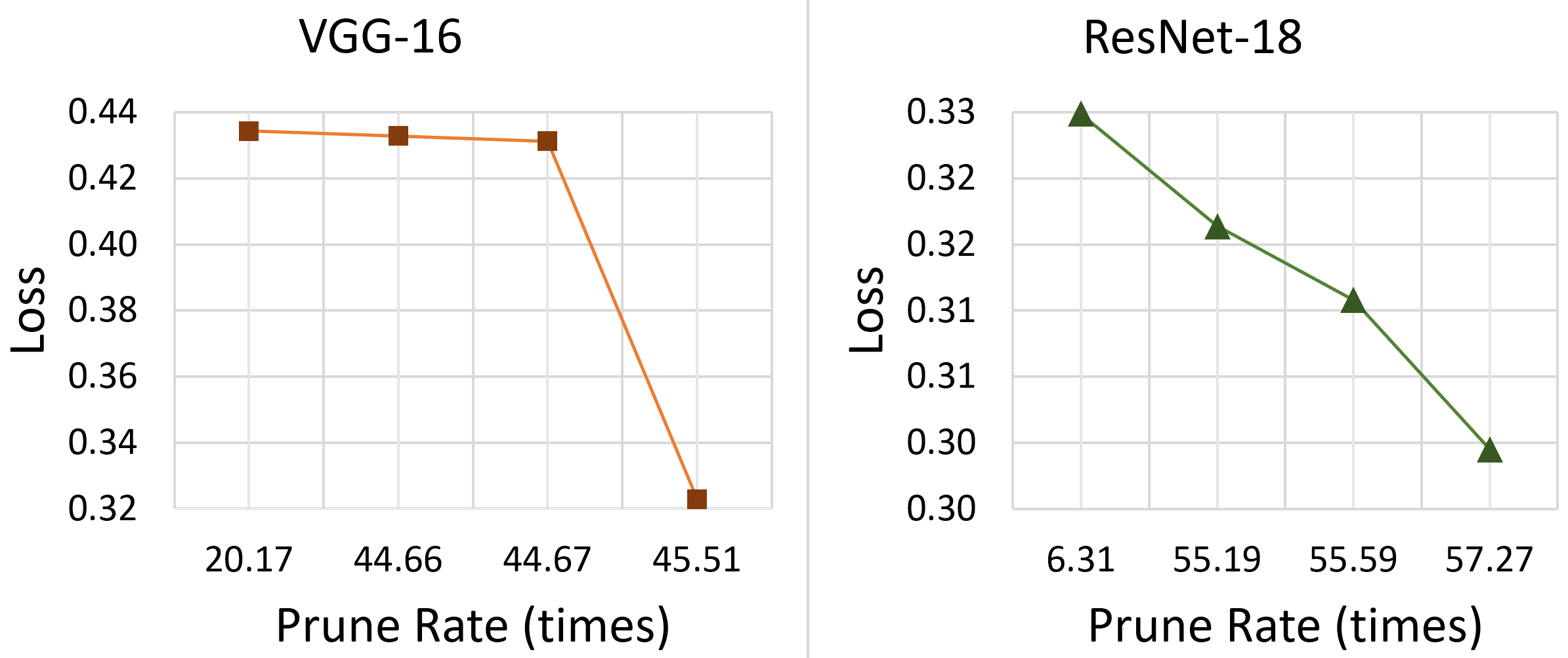} 
     \vspace{-3mm}
     \caption{Effect of removing redundant weights and unused paths. (dataset: CIFAR-10; Accuracy: VGG-16-93.36\%, ResNet-18-93.79\%)}
     \label{fig:purify_loss}
\end{figure}

As shown in Table~\ref{table:results}, we show groups of different prune ratios and 8-bits quantization with accuracies on each network structure. 
Figure \ref{fig:purify_loss} proves our previous arguments that ADMM's non-optimality exists in a structured pruned model. P-RM can further optimize the loss function. Please note all of the results are based on non-retraining process. Below are some results highlights on different dataset with different network structures.

\emph{\textbf{MNIST.}} With LeNet-5 network, comparing to original accuracy (99.17\%), our proposed P-RM framework achieve 231.82$\times$ compression with minor accuracy loss while other state-of-art compression ratios are lossless. And no accuracy losses are observed after quantization on 40$\times$ and 88$\times$ models and only 0.4\% accuracy drop on 231.82$\times$ model. On the other hand, Group Scissor~\cite{wang2017group} only has 4.16$\times$ compression rate.

\emph{\textbf{CIFAR-10.}} Convnet structure are relative shallow so ADMM reaches a relative optimal local minimum, so post-processing is not necessary. But we still outperform Group Scissor~\cite{wang2017group} in accuracy (84.55\% to 82.09\%) when compression rate is same (2.35$\times$). For larger networks, when a minor accuracy loss is allowed, our proposed P-RM method improves the prune ratio to 50.02$\times$ and 59.84$\times$ on VGG-16 and ResNet-18 respectively, and no obvious accuracy loss after quantization on pruned models.

\emph{\textbf{ImageNet.}} AlexNet model outperform SSL~\cite{wen2016learning} both in compression rate (4.69$\times$ to 1.40$\times$) and network accuracy (81.76\% to 80.40\%), with or without P-RM. Our ResNet-18 and ResNet-50 models also achieve unprecedented 3.33$\times$ with 88.36\% accuracy and 2.70$\times$ with 92.27\% respectively. No accuracy losses are observed after quantization on pruned ResNet-18/50 models and around 1\% accuracy loss on 5.13$\times$ compressed AlexNet model.

Table~\ref{table:power_and_area} shows our highlighted memristor crossbar power and area comparisons of ResNet-18 and VGG-16 models. By using our proposed P-RM method, the area and power of the $5.83\times$ $(15.14\times)$ ResNet-18 model is reduced from 0.235$mm^2$ (0.117$mm^2$) and 3.359$W$ (1.622$W$) to 0.042$mm^2$ (0.041$mm^2$) and 0.585$W$ (0.556$W$), without any accuracy loss. For VGG-16 $20.16\times$ model, after using our P-RM method, the area and power is reduced from 0.113$mm^2$ and 1.611$W$ to 0.056$mm^2$ (0.053$mm^2$) and 0.824$W$ (0.754$W$), where the compression ratio is achieved 44.67$\times$ (50.02$\times$) with 0\% (0.63\%) accuracy degradation.


\begin{table}[t]
  \caption{Area/power comparison between models with and without P-RM on ResNet-18 and VGG-16 on CIFAR-10}
  \label{table:power_and_area}
  \vspace{-0mm}
  \includegraphics[width=\linewidth]{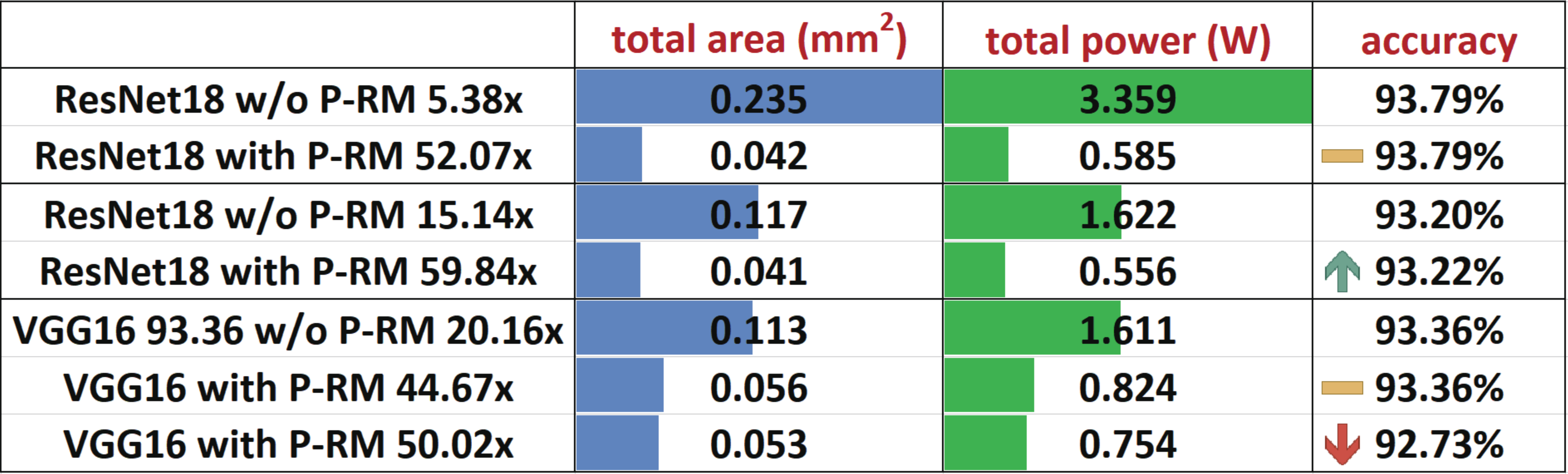}
\end{table}

\section{Conclusion}

In this paper, we designed an unified memristor-based DNN framework which is tiny in overall hardware footprint and accurate in test performance. We incorporate ADMM in weight structured pruning and quantization to reduce model size in order to fit our designed tiny framework. We find the non-optimality of the ADMM solution and design \textit{Network Purification} and \textit{Unused Path Removal} in our software-hardware co-optimization framework, which achieve better results comparing to Gourp Scissor~\cite{wang2017group} and SSL~\cite{wen2016learning}. On AlexNet, VGG-16 and ResNet-18/50, after structured weight pruning and 8-bit quantization, model size, power and area are significant reduced with negligible accuracy loss.


\scriptsize

\end{document}